# U-Slot Patch Antenna Principle and Design Methodology Using Characteristic Mode Analysis and Coupled Mode Theory

John J. Borchardt, *Member, IEEE* and Tyler C. LaPointe

*Abstract*—Patch antennas incorporating a U-shaped slot are well-known to have relatively large (about 30%) impedance bandwidths. This work uses Characteristic Mode Analysis to explain the impedance behavior of a classic U-slot patch geometry in terms of Coupled Mode Theory and shows the relevant modes are in-phase and anti-phase coupled modes whose resonant frequencies are governed by Coupled Mode Theory. Additional analysis shows that one uncoupled resonator is the conventional $TM_{01}$ patch mode and the other is a lumped LC resonator involving the slot and the probe. An equivalent circuit model for the antenna is given wherein element values are extracted from Characteristic Mode Analysis data and which explicitly demonstrates coupling between these two resonators. The circuit model approximately reproduces the impedance locus of the driven simulation. A design methodology based on Coupled Mode Theory and guided by Characteristic Mode Analysis is presented that allows wideband U-slot patch geometries to be designed quickly and efficiently. The methodology is illustrated through example.

## I. INTRODUCTION

Huynh and Lee [1] showed the addition of a U-shaped slot significantly increased the otherwise narrow impedance bandwidth (BW) of a probe-fed microstrip patch antenna on a low permittivity (foam) substrate. It was hypothesized at the time that the increased impedance bandwidth was due to the existence of two resonances—that of the patch and that of the U-shaped slot. Subsequent investigation [2], [3], [4] found the achievable pattern bandwidth of U-slot patches on low permittivity substrates was around 30%. In [5], workers used full-wave simulation to develop a U-slot patch design on an $\epsilon_r = 2.33$ substrate with 25% impedance bandwidth.

Researchers subsequently reported observations of, and *empirical* design algorithms for, the U-slot patch. One study [6] gave qualitative guidelines as to how the impedance locus behaved in response to dimensional changes. Another study [7] found *empirical* relations between design dimensions and the frequencies of the reflection coefficient magnitude minima. In [8], investigators used numerical studies to characterize the *empirical* response of the impedance locus to dimensional changes and gave an algorithm that yields initial design dimensions. Another *empirical* study [9] observed that the ratios of acceptable design dimensions were substantially constant with changes in substrate permittivity and gave formulas for initial U-slot patch dimensions. Some success modeling the U-slot patch with an equivalent circuit was reported in [10].

Characteristic Mode Analysis (CMA) has been applied to U-shaped slots and U-slot patches in the past; here we make important distinctions between these works and the present study. For example, in [11] researchers applied CMA to circularly polarized U-slot patches, however, these devices are mostly unrelated to the wideband, linear-polarized U-slot patches of this work. Studies [12], [13] and [14] concerned CMA mode tracking algorithms and gave examples for U-shaped slots in ground planes or plates; however, these structures are not patch antennas per se. In [15], CMA was applied to a U-slot patch *without a feed probe*—in contrast to

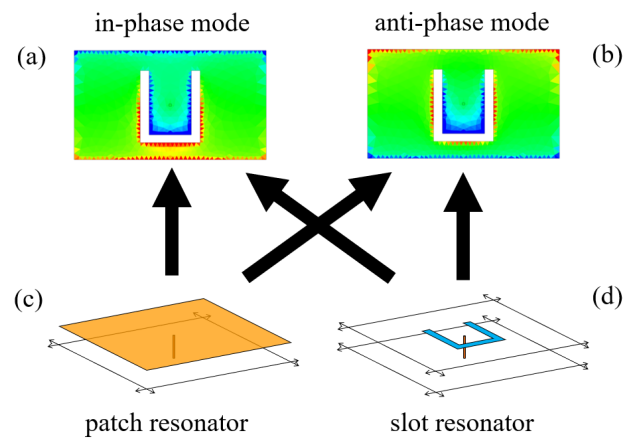

*Figure 1.* CMA charge distributions of (a) in-phase and (b) anti-phase coupled modes; each comprises (c) uncoupled patch and (d) uncoupled slot resonators described in Section III. Section IV shows these are coupled via a mutual inductance related to the U-slot width.

---

this work; no design guidelines or fundamental operational principles were given there. In [16], CMA was used to examine the *empirical* U-slot patch design methods of [8] and [9] and gave a third *empirical* design methodology based on a combination of the two. In [17], CMA was used to examine the effect of probe location on U-slot patch modes and impedance. CMA-based optimization of the slot shape and probe location in U-slot patches was presented in [18]. Neither [17] nor [18] addresses the U-slot patch initial design process or gives a fundamental operational mechanism.

Although any antenna may be designed purely via full-wave simulation with numerical optimization, first-principles models are invaluable both for generating good designs quickly as well as understanding the limitations and possibilities of device performance. Unfortunately, a comprehensive, first-principles explanation and quantitative design method based on such for wide-band, linearly polarized U-slot patches is uncommon in the literature. This work uses Characteristic Mode Analysis (CMA) and Coupled Mode Theory (CMT) to characterize the U-slot patch in a new way and develops a design methodology based directly on the given operational principles, extending earlier analysis [19], [20] that gave evidence that the two resonances of the U-slot patch are related to CMT, as illustrated in Fig. 1. The paper contributes to the understanding and design of U-slot patches by: 1) demonstrating that the classic U-slot patch [1] is governed by CMT, 2) clearly identifying both resonators of the U-slot patch, 3) developing a successful equivalent circuit that explicitly shows coupling between two resonators, 4) giving a bandwidth-optimal procedure for determining coupling, and 5) presenting a design methodology based on the operational principle. With the methodology presented, U-slot patches may be designed quickly and efficiently. Readers with no background in CMA are referred to the introductory material in the Appendix and the references cited there.

## II. MODAL ANALYSIS

### A. Characteristic Mode Analysis

FEKO, a method of moments (MoM) solver with CMA [21], is used to analyze the U-slot patch geometry of [1] shown in Fig. 2. Conductors are modeled as ideal and thus the calculated radiation efficiency is 100%; small losses may be treated as a perturbation. The probe is modeled as a cuboid with cross-section $2.7\text{ mm} \times 2.7\text{ mm}$ (equal to the round probe area in [1]). Modeling the probe is essential because it serves as the inductance in the *uncoupled* slot resonator, as described in Section III. The CMA eigenvalues $\lambda_n$ are shown in Fig. 3(a). Modes 1 and 3 (numbering is arbitrary) are resonant ($\lambda_n = 0$) near 0.80 and 1.05 GHz, respectively. Fig. 3(b) shows the modal weighting coefficients (using (16) of the Appendix) due to a 1V gap source at the base of the probe and demonstrates modes 1 and 3 are the only strongly excited modes.

Fig. 4 shows the full-wave driven admittance of the U-slot patch; the 6dB return loss BW of the driven full-wave locus is $0.78 - 1.09$ GHz and the center frequency, $f_0$, is 940 MHz. The admittance of modes 1 and 3 at the gap source is also shown in Fig. 4. According to (15) of the Appendix, the total admittance is the parallel combination of individual modal admittances, and this is also plotted for modes 1 and 3 in Fig. 4. This locus differs from the driven full-wave locus by only a small shunt capacitance attributed to sub-resonant higher order modes, and demonstrates that modes 1 and 3 are the only modes relevant to U-slot patch operation.

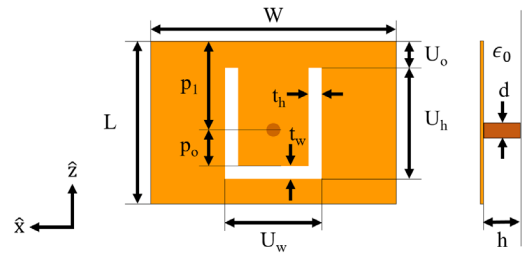

Figure 2. U-slot patch geometry of [1] where $W$=220 mm, $L$=124 mm, $h$=26.9 mm, $U_w$=68.6 mm, $U_h$=82.2 mm, $U_0$=22.9 mm, $t_h$=10.2 mm, $t_w$=8.89 mm, $d$=3.05 mm, and $p_0$=33.9 mm. The coordinate system origin is at the base of the probe.

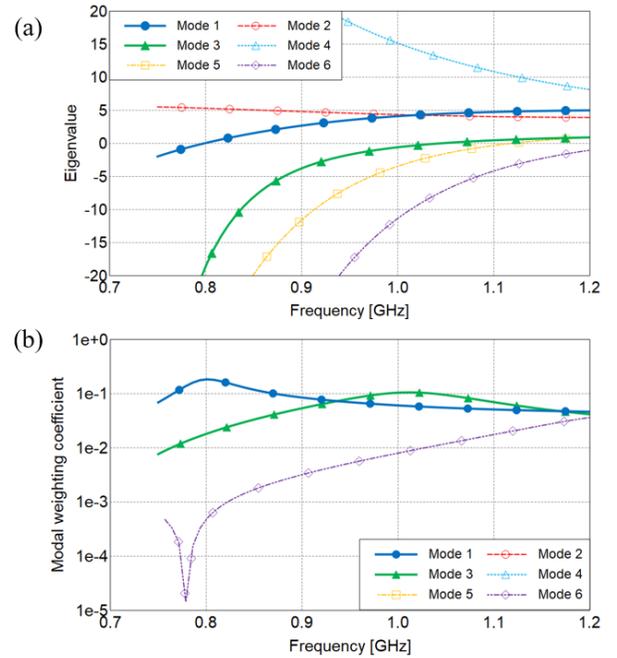

Figure 3. (a) U-slot patch CMA eigenvalues; (b) the modal weighting coefficients $\alpha_n$ show modes 1 & 3 are the only strongly excited modes (modes 2, 4 & 5 have $|\alpha_n| <$ 1e-5). Mode 6 is a perturbed $TM_{20}$ mode that is weakly excited within the impedance bandwidth but responsible for cross-polarized radiation at high frequencies as discussed in [4].

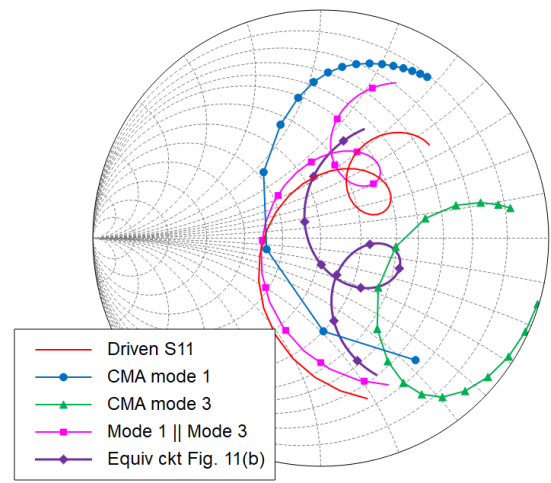

Figure 4. The parallel combination of CMA modes 1 & 3 closely replicates the driven impedance locus of the U-slot patch, demonstrating they are the only important modes.

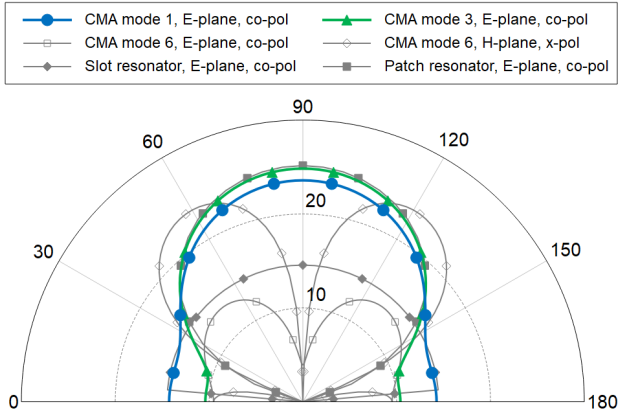

*Figure 5. Modal far-fields (dBV/m at $r = 1$ meter); CMA modes 1 & 3 have broadside, co-polarized patterns (i.e., θ-polarized in the y-z plane). As discussed in Section II, CMA mode 6 is predominantly cross-polarized in the H-plane (ϕ-polarized in the x-y plane). As discussed in Section III, the uncoupled slot resonator far-fields are about 10dB less than those of the uncoupled patch.*

Fig. 5 shows E-plane far-field patterns for modes 1 and 3. Both have broadside, co-polarized radiation patterns, like that of the conventional $TM_{01}$ patch mode, resulting in a stable radiation pattern throughout the entire impedance bandwidth.

### B. Coupled Mode Theory

CMT is relevant to a wide variety of physical phenomenon [22] and summarized in [23]; it states a system of two coupled resonators can be analyzed as the superposition of two modes with lower- and higher-frequencies wherein the resonators move in-phase and anti-phase, respectively.

The coupled mode frequencies, $\omega_+$ and $\omega_-$, are related to the uncoupled mode frequencies, $\omega_1$ and $\omega_2$, by [24]:

$$\omega_\pm = \omega_0 \pm \sqrt{\left(\frac{\omega_2 - \omega_1}{2}\right)^2 + |K|^2} \qquad (1)$$

where $\omega_0 = (\omega_2 + \omega_1)/2$ and $K$ is an un-normalized coupling coefficient. Given $\omega_1 = \omega_2$, a normalized coupling coefficient may be calculated via [25]:

$$\kappa = \frac{\omega_+^2 - \omega_-^2}{\omega_+^2 + \omega_-^2}. \qquad (2)$$

Inserting (1) with $\omega_1 = \omega_2$ into (2) four times gives $\kappa = (2\omega_0 K)/(\omega_0^2 + K^2)$. Given $K^2 \ll \omega_0^2$, we have:

$$K \sim \omega_0 \kappa / 2. \qquad (3)$$

The current, charge and electric field distributions for modes 1 and 3 near their respective resonant frequencies are shown in Fig. 6. Charge accumulation is visible at the edges of the patch and the center of the slot in Fig. 6(c) and 6(d), however, the spatial orientation of the two differs between modes. For mode 1, the patch and slot charge distributions are in phase; for mode 3, they are anti-phase. This suggests that CMT is relevant to the U-slot patch.

Further evidence of the role of CMT in the U-slot patch is found in how the CMA resonances respond to changes in coupling coefficient. According to (1), greater coupling yields a larger difference between the coupled mode resonant frequencies. We propose that $U_w/W$ approximates the fraction of the unperturbed $TM_{01}$ patch mode current intercepted by the slot and thus controls the coupling. Accordingly, the difference in resonant frequencies should increase with greater $U_w$. This behavior is demonstrated in Fig. 7 for the geometry of [1]; here, only $U_w$ is varied while all other dimensions remain constant. From Fig. 7, a first-order approximation for $\kappa$ (calculated via (2) using the CMA resonant frequencies) is:

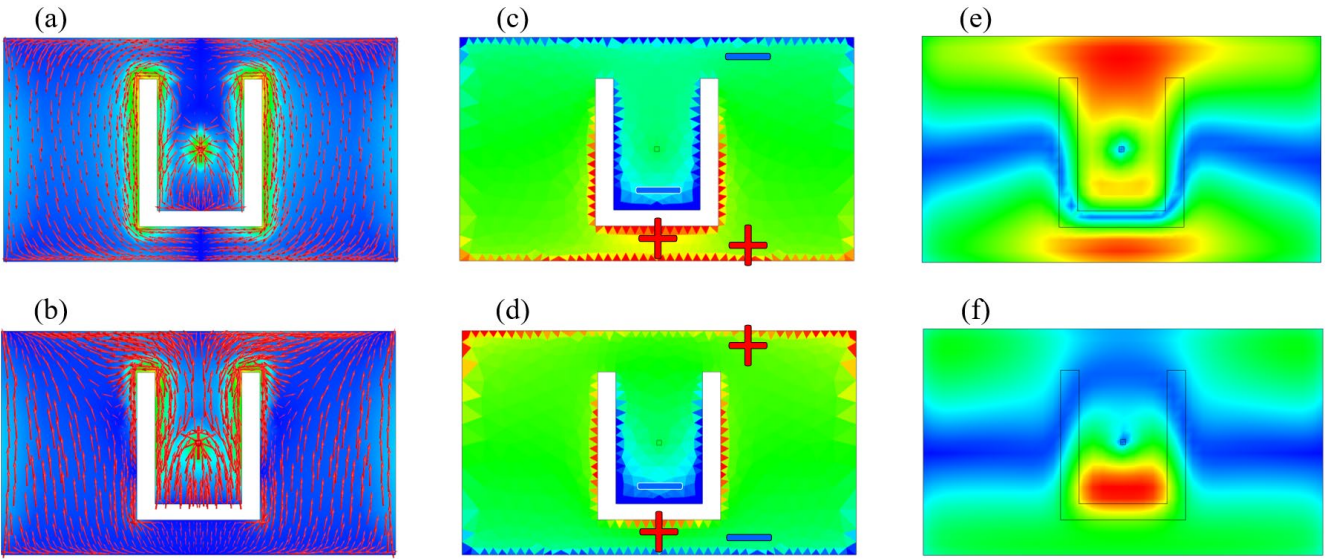

*Figure 6. Normalized current distributions (a) & (b); normalized charge distributions (c) & (d); normalized electric field magnitude distributions (e) & (f) normal to the plane $y = h/2$ for CMA mode 1: (a), (c) & (e) and CMA mode 3: (b), (d) & (f). The charge distributions show the distinctive in-phase and anti-phase relationships characteristic of Coupled Mode Theory.*

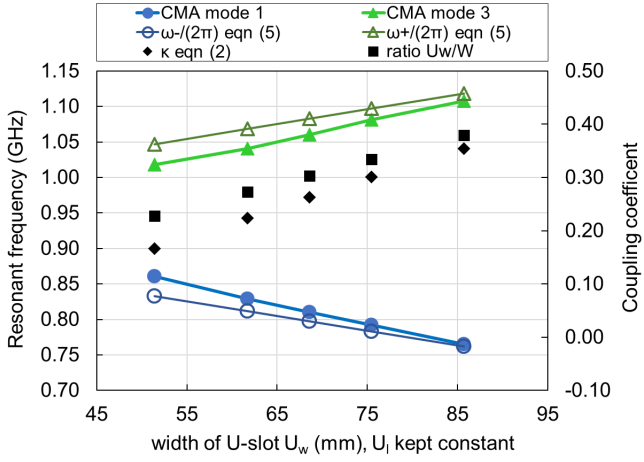

*Figure 7. Left axis (solid lines): Increasing the U-slot width (while keeping the U-slot total length and all other dimensions constant) increases the coupling coefficient κ and thus the difference in resonant frequencies according to (1). Right axis (♦ and ■ data points): the coupling coefficient calcualted via (2) is approximated by the ratio $U_w/W$. The CMT-derived relation (5) accurrately predicts the CMA coupled resonant frequencies to within a few percent.*

$$\kappa \sim U_w/W. \quad (4)$$

Combining (3), (4) and (1) with $\omega_1 = \omega_2$ then yields:

$$\omega_\pm = \omega_0\,(1 \pm U_w/(2W)) \quad (5)$$

where $\omega_0 = 2\pi f_0$ is the center frequency. *As shown in Fig. 7, this CMT-derived relation accurately predicts the CMA resonant frequencies to within a few percent and demonstrates CMT underlies the operation of the classic U-slot patch described in [1].*

Like this work, [26] observed that the fields of the U-shaped slot and patch edges have in-phase and anti-phase relationships depending on frequency; however, that study did not invoke Coupled Mode Theory. Reference [27] argued the second resonance (aside from the $TM_{01}$ patch mode) is a perturbed $TM_{20}$ mode; this work gives evidence that this mode (CMA mode 6) is only weakly excited (see Fig. 3(b)) and is responsible for cross-polarized radiation discussed in [4] (see Fig. 5).

### III. Uncoupled Resonators

CMA of the patch *with no slot* shows that the $TM_{01}$ mode is resonant at 0.94 GHz—near the impedance bandwidth center frequency $f_0$. Given this, (1) implies the other coupled resonance will also be near 0.94 GHz. However, CMA of the U-slot in a single conducting plane yields a mode resonant at 0.64 GHz (where $U_l \sim \lambda/2$). We instead represent the uncoupled slot resonator as the U-slot in one of two parallel, infinite conducting planes, separated by the patch dielectric substrate, and connected by the feed probe, as shown in Fig. 8. This is equivalent to the full U-slot patch geometry as $W$ and $L$ are increased to infinity. The geometry supports electric current J on the probe and magnetic current M on the slot; the infinite ground planes are accounted for via the Green's function in the MoM code used [21]. For brevity, we call the Fig. 8 geometry the "uncoupled slot resonator" although it equally involves the probe and ground plane, as discussed below.

We note that the Fig. 8 geometry differs from the U-slot in a single conducting plane and therefore has different properties. We may deduce some of these properties via CMA of the geometry with either a perfect magnetic conductor (PMC, $\hat{n} \circ \vec{E} = 0$) or perfect electric conductor (PEC, $\hat{n} \times \vec{E} = 0$) boundary on the x-z plane. Two modes of interest result, with CMA mode 4 (resonant near $f_4 = 0.70$ GHz) satisfying the PMC boundary condition and CMA mode 2 (resonant near $f_2 = $ 0.91 GHz), satisfying the PEC boundary condition (see Fig. 9). The vertical electric fields of Fig. 9(d) indicate there is capacitive coupling, $C_p$, between the slot edge and its ground plane image for the PEC case only; $C_p$ is in parallel with the probe inductance $L_p$ (see Fig. 9(b)). Moreover, CMA modes 4 and 2 can be interpreted as in-phase and anti-phase coupled modes, respectively; the distinction is made according to the slot magnetic current orientation with respect to its image, as dictated by the x-z plane boundary condition.

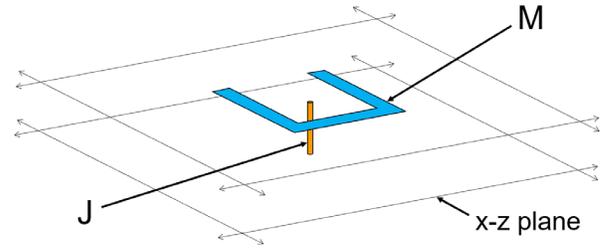

*Figure 8. The uncoupled slot resonator is the U-slot geometry of Fig. 2 with W and L increased to infinity; this structure has a mode resonant near $f_0$. Electric currents J are supported on the probe and magnetic currents M on the slot. Modeling the probe is essential because it provides part of the inductance of the resonator.*

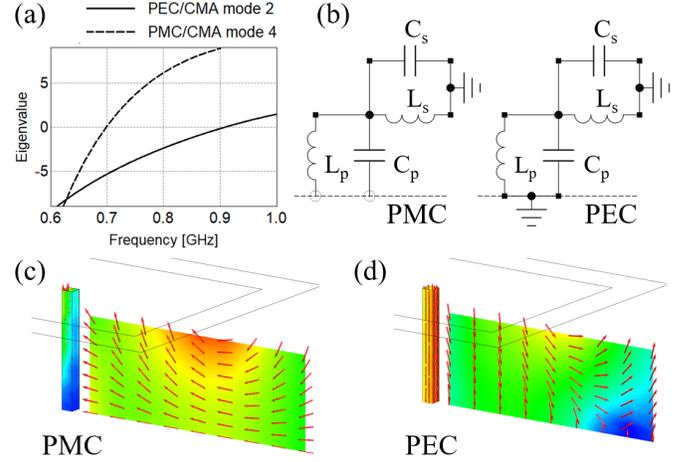

*Figure 9. (a) CMA eigenvalues of, and (b) equivalent circuit for, the geometry of Fig. 8 with either PMC or PEC boundary on the x-z plane. (c) CMA mode 4 and (d) CMA mode 2 electric fields. Vertical electric field lines in (d) indicate a parasitic capacitance $C_p$ couples the slot to its ground plane image (not shown); there are no such lines in (c).*

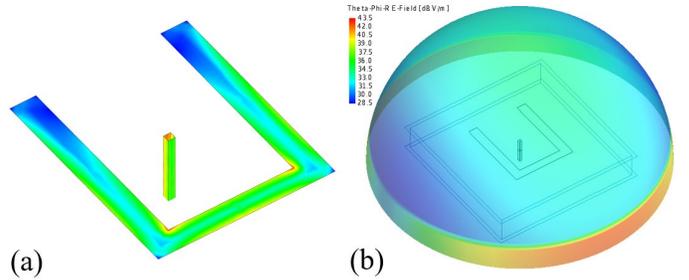

*Figure 10. (a) Uncoupled slot CMA mode 2 electric and magnetic current magnitudes show a uniform probe electric current and a cosine variation of the slot magnetic current; (b) modal near-fields show that the probe primarily excites the parallel plate waveguide formed by the two infinite ground planes and radiates comparatively little power.*

The equivalent circuits of Fig. 9(b) correspond to the two boundary conditions imposed. The slot resonance is represented by $L_s$ and $C_s$; its extremity is grounded because the slot is within an infinite conducting plane. With a PMC boundary, $C_p$ and $L_p$ are open-circuited; thus, the resonance is $1/\sqrt{L_s C_s}$. Despite the nearby PMC plane, the resonance is close to that of the U-slot in a single conducting plane. With a PEC boundary (grounded due to symmetry), all components are in parallel and thus the resonance is $1/\sqrt{L_{eff} C_{eff}}$ where $L_{eff} = (L_p L_s)/(L_p + L_s)$ and $C_{eff} = C_s + C_p$. We now estimate the circuit element values. Using the magnetic frill model of [28], we calculate $L_p$ = 12.2 nH at 0.805 GHz (midway between the mode 4 and 2 resonances). We estimate the static slot capacitance using the method of [29] as $C_s \sim 3.8$ pF and deduce $L_s \sim 1/((2\pi f_2)^2 C_s) = 13.2$ nH. We estimate $C_p \sim 1.2$ pF via the parallel plate capacitor formula (using the conductor area between the probe and the horizontal portion of the U-slot). With these values, the Fig. 9(b) circuit resonances are 0.70 GHz and 0.89 GHz—close to those of CMA modes 4 and 2.

CMA mode 2 (with PEC boundary) is the mode relevant to the U-slot patch because in the geometry of Fig. 2, the x-z plane is electrically conducting. We note the resonance is determined both by the slot resonance as well as $L_p$ and $C_p$. Moreover, the mode 2 resonance fits well within the CMT framework established in Section II (i.e., $\omega_1 \sim \omega_2$ in (1)).

The CMA mode 2 current magnitudes near resonance are shown in Fig. 10(a). Again, this mode behaves as a lumped LC resonator among $L_s$, $L_p$, $C_s$ and $C_p$; e.g., increasing the probe diameter $d$ and slot thickness $t_w$ and $t_h$ reduces $L_p$ and $C_s$, respectively, thereby increasing the uncoupled slot resonator frequency $f_{slot}$. However, when the structure is fed at the probe base, these dimensions do not appear to strongly affect the overall resonant conductance, $G_{0,slot} \sim \sqrt{C/L}/Q$, where $Q$ is the quality factor [30]. The resonant conductance is ultimately important for obtaining the desired impedance locus as discussed in Section V. $G_{0,\,slot}$ and $f_{slot}$ are both more strongly influenced by $U_l$, $h$, $\epsilon$, and $p_0$. E.g., increasing $h$ increases $L$, lowering $f_{slot}$ and decreasing $G_{0,\,slot}$. Increasing $\epsilon$ increases $C$, lowering $f_{slot}$ and increasing $G_{0,\,slot}$.

As seen in Fig 10(b), the CMA mode 2 near-fields indicate the feed probe strongly excites the parallel plate waveguide formed by the infinite ground planes; thus, the structure radiates comparatively little power. At resonance, the maximum modal far-field amplitude (normalized to $r = 1$ meter) is 14.5dBV/m at broadside; the corresponding amplitude for the uncoupled patch resonator is 25.1dBV/m—similar to that of CMA modes 1 and 3 of the full U-slot patch geometry, as shown in Fig. 5. This is evidence that radiation from the full U-slot patch structure is due predominantly to the patch edges; this is reasonable given that the total length of the patch edges, $2W$, is about 6.5 times the slot width, $U_w$, and that the patch height, $h$, is about 3 times the slot thickness, $t_w$. Thus, the slot magnetic current which opposes that of the patch edges in the anti-phase mode of the full U-slot patch geometry does not significantly impact the radiation pattern or the directivity of the anti-phase mode, as shown in Fig. 5.

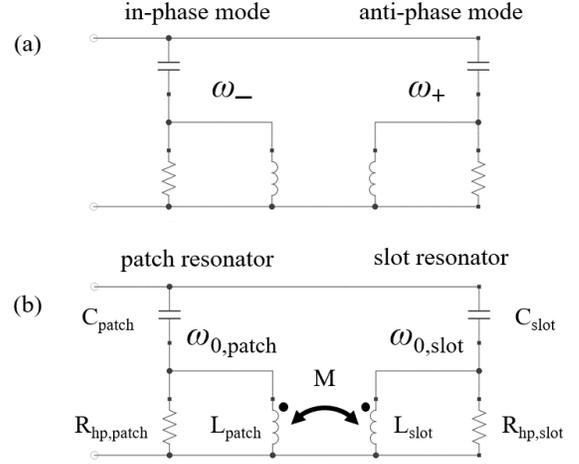

Figure 11. Equivalent circuit models for the U-slot patch: (a) broadband equivalent circuit based directly on the in-phase and anti-phase CMA modal admittances (which are orthogonal and have no coupling); (b) circuit model of uncoupled patch and slot resonators where coupling is explicitly shown through a mutual inductance.

IV. EQUIVALENT CIRCUIT MODEL

The impedance of a characteristic mode may be modeled as a first-order high-pass RLC circuit [31]—although other representations are possible. The parallel combination of two such circuits, seen in Fig. 11(a), captures the net impedance behavior of a geometry where two characteristic modes are relevant, as indicated by (15). Here, there is no inter-modal coupling due to orthogonality of the characteristic modes.

The circuit model of Fig. 11(b) qualitatively reproduces the impedance locus of the U-slot patch. Note $L_{slot}$ and $C_{slot}$ are different than $L_s$ and $C_s$ discussed in Section III. Here, the uncoupled patch and slot resonators are explicitly coupled through a mutual inductance. This is justified given that a slot voltage is proportional to the time-derivative of the current it intercepts; in this case, the U-slot can be thought of as intercepting the $TM_{01}$ mode patch current. The mutual inductance is $M = k\sqrt{L_{slot}L_{patch}}$ where $k = \kappa$; given the CMA mode 1 and 3 resonant frequencies, (2) yields $\kappa = 0.26$.

For each resonator, RLC values may be calculated from the *uncoupled* resonator CMA data ($\omega_0$, $G_0$ and $Q$) via circuit analysis:

$$Y_c = G_0/Q' \qquad R_{hp} = G_0/Y_c^2$$
$$C = Y_c/\omega_0 \qquad L = 1/(Y_c\,\omega_0) \qquad (6)$$

where $Q' = \sqrt{Q_{slot}Q_{patch}}$. The uncoupled resonator CMA data and resulting circuit values from (6) are given in Table I.

Despite the extreme RLC values of the patch resonator, the agreement between the Fig. 11(b) circuit model and driven impedance loci in Fig. 4 is fair, with a modest 10% frequency shift between reflection coefficient magnitudes, as shown in Fig. 12. The RLC values in Table I imply that the slot resonator plays an impedance matching role in the U-slot patch (recall the slot itself does not radiate strongly). When coupled, the slot and patch resonators together produce the stagger-tuned in-phase and anti-phase coupled modes that yield broad impedance bandwidth.

TABLE I
UNCOUPLED RESONATOR CMA DATA AND RESULTING
EQUIVALENT CIRCUIT VALUES FOR HUYNH AND LEE DESIGN [1]

| Patch resonator | | Slot resonator | |
|---|---|---|---|
| $f_{\text{patch}}$ | 946 MHz | $f_{\text{slot}}$ | 912 MHz |
| $G_{0,\text{patch}}$ | 1.0 µS | $G_{0,\text{slot}}$ | 34 mS |
| $Q_{\text{patch}}$ | 4.5 | $Q_{\text{slot}}$ | 8.9 |
| $Q'$ | 6.3 | $Q'$ | 6.3 |
| $R_{\text{hp,patch}}$ | 40 MΩ | $R_{\text{hp,slot}}$ | 1.18 kΩ |
| $L_{\text{patch}}$ | 1.06 mH | $L_{\text{slot}}$ | 32.5 nH |
| $C_{\text{patch}}$ | 0.0266 fF | $C_{\text{slot}}$ | 0.938 pF |
| $k$ | 0.26 | $k$ | 0.26 |

TABLE II
BANDWIDTH-OPTIMAL DUAL RESONATOR PARAMETERS

| Return loss (dB) | $y_{\text{opt}}$ | $G_{\text{opt}}/Y_0$ | $BW_x$ |
|---|---|---|---|
| 6 | 3.62 | 2.38 | 8.46 |
| 8 | 2.76 | 1.85 | 6.23 |
| 10 | 2.25 | 1.56 | 4.86 |
| 12 | 1.92 | 1.41 | 3.94 |
| 16 | 1.53 | 1.22 | 2.75 |
| 20 | 1.32 | 1.12 | 2.03 |

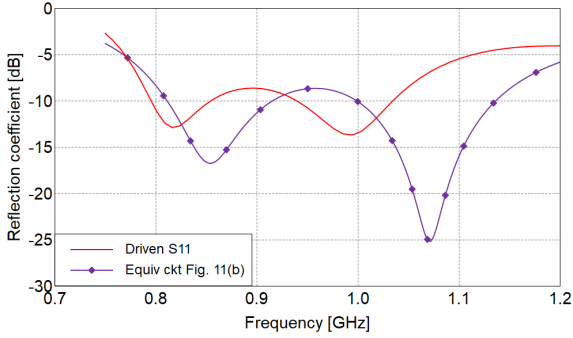

Figure 12. Reflection coefficient magnitude (in 50 Ω) of full-wave driven geometry [1] and the equivalent circuit of Fig. 11(b).

It is difficult to directly compare the modes of the full geometry to those of the uncoupled patch and uncoupled slot resonators because they have different support. However, the success of the equivalent circuit model implies that the modes of the full geometry are indeed coupled forms of the uncoupled patch and slot resonator modes; recall that the circuit element values of Fig. 11(b) are extracted from CMA data of each separate and uncoupled resonator and that the coupling coefficient is derived from the full geometry resonant frequencies via (2).

We also note that the equivalent circuit of Fig. 11(b) consists of two resonators coupled by a mutual inductance—a classic example often used to illustrate CMT. Moreover, eigen-analysis of this circuit shows that it supports modes wherein the inductor currents are in-phase and anti-phase—the distinct signature of CMT. It is reasonable to conclude that if CMT governs an equivalent circuit that accurately models the U-slot patch, then CMT also governs the U-slot patch.

V. DESIGN METHODOLOGY

The design methodology suggests an initial structure and systematically refines it using simulation; it consists of three steps. First, the desired coupling coefficient is established.

Second, uncoupled patch and slot resonators with approximately equal resonant frequencies are designed using CMA. Third, the two geometries are combined into a full U-slot patch structure and the full-wave impedance is calculated. If necessary, the geometry may be refined with a few simple guidelines to yield an improved impedance locus.

A. Bandwidth-Optimal Stagger-Tuned Resonances

The admittance of a parallel combination of two stagger-tuned series resonators of resonant frequency $\omega_1$ and $\omega_2$ with equal $Q$ and resonant conductance $G_0$ in terms of a normalized frequency $x$ and a resonant frequency separation $y$ is [32]:

$$Y = 2G_0 \frac{(1+jx)}{(1+j(x+y))(1+j(x-y))}, \quad (7)$$

where

$$x = 2Q\frac{\omega - \omega_0}{\sqrt{\omega_1 \omega_2}}, \qquad y = Q\frac{\omega_2 - \omega_1}{\sqrt{\omega_1 \omega_2}}, \quad (8)$$

and $\omega_0 = (\omega_1 + \omega_2)/2 = 2\pi f_0$. For our purposes, the resonances $\omega_1$ and $\omega_2$ refer to those of the *coupled* system (i.e., those of the full U-slot patch geometry).

A numerical optimization of (7) for greatest 10 dB return loss (RL) bandwidth in a system impedance $Z_0 = 50\,\Omega = 1/Y_0$ yields $y_{\text{opt}} = 2.25$ and $G_{\text{opt}} = 31$ mS with normalized bandwidth $BW_x = 4.86$. The resulting admittance locus is shown in Fig. 13 and its corresponding reflection coefficient magnitude is plotted in Fig. 14. A misconception regarding such frequency responses is that the resonant frequencies correspond to the minima of the reflection coefficient magnitude. This is not the case in general; the resonances occur at normalized frequencies $x = \pm y$ (e.g., setting $x = \pm y$ yields $\omega = \omega_{2,1}$). Bandwidth-optimal values of $y$ and $G_0$ generated by numerical optimization are in Table II.

With $y_{\text{opt}}$ determined, the in-phase and anti-phase coupled resonant frequencies $f_\pm = \omega_\pm/(2\pi)$ are:

$$f_\pm = f_0\left(1 \pm \frac{y_{\text{opt}}}{2Q}\right). \quad (9)$$

Now we must determine the $Q$ required to support the desired unnormalized impedance bandwidth $BW$. From Fig. 14, a simple approximation is:

$$BW \sim f_+ - f_-. \quad (10)$$

Combining (9), (10) and (2) with a Taylor series approximation yields:

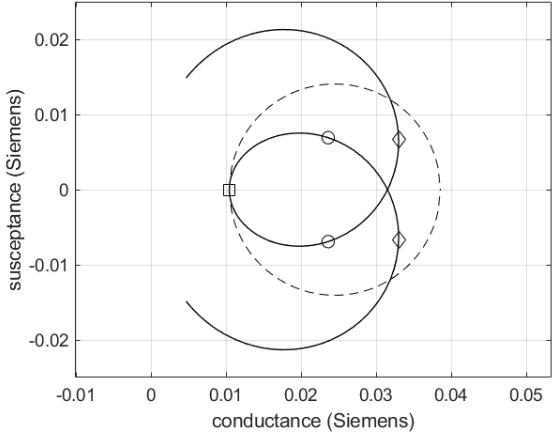

Figure 13. Maximum BW admittance locus (7) for a RL of 10 dB in $Z_0 = 50$ Ω. The 10 dB RL circle is shown as a dashed line. Resonant frequencies, reflection coefficient minima, and the center frequency are marked by '◊', '○' and '□', respectively.

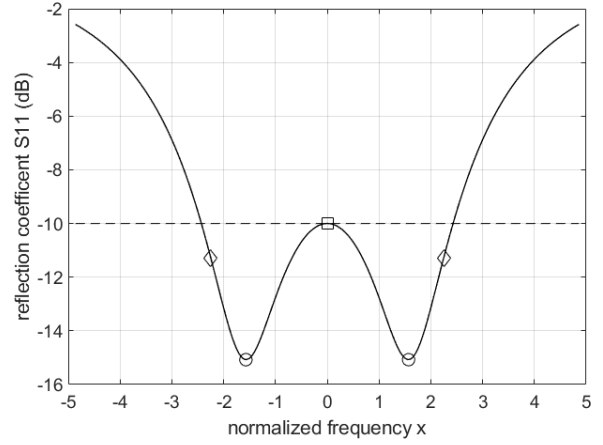

Figure 14. Reflection coefficient versus normalized frequency $x$ of the admittance locus of Fig. 10; 10 dB RL is indicated by a dashed line. Resonant frequencies, reflection coefficient minima, and the center frequency are marked by '◊', '○' and '□', respectively.

$$\frac{BW}{f_0} \sim \frac{y_{opt}}{Q} \sim \kappa. \quad (11)$$

Patch parameters $L$, $W$, $h$ and $\epsilon$ can now be selected to achieve a radiation $Q$ in accordance with (11).

Recall that $G_{opt}$ refers to the *coupled* modes; we seek to establish $G_0$ of the *uncoupled* modes. Because the probe is close to the center of the patch, $G_{0,patch}$ will be small. Given this, a numerical study of the Fig. 11(b) circuit shows the proper $G_{0,slot}$ will be about $1.5 \times G_{opt} \sim 40 - 50$ mS for $Z_0 = 50$ Ω and 10 dB RL. After coupling, the coupled mode resonant conductances will be close to $G_{opt}$.

*B. Designing the Uncoupled Resonators*

CMA of the uncoupled patch resonator (with no U-slot) is performed, and the geometry set so that the TM$_{01}$ mode resonant frequency $f_{patch} = f_0$. The $Q$ must be less than that dictated by (11). Good initial parameters are $W = 2L$ and $p_1 = L/2$. At this point, the probe near the patch center has little effect on $f_{patch}$ and $Q_{patch}$.

Design of the uncoupled slot resonator (shown in Fig. 8) is less straightforward. We seek $f_{slot} = f_0$ and $G_{0,slot} = 40 - 50$ mS. The probe presence and location affect $f_{slot}$, increasing it above where $U_l = \lambda/2$. Design guidelines are:

- $U_w/W$ should be approximately equal to $\kappa$.
- $U_l$ has the greatest influence on $f_{slot}$; start with $U_l = \lambda/2$; increasing $U_l$ lowers $f_{slot}$ and increases $G_{0,slot}$.
- Increasing $p_o$ lowers $f_{slot}$ and increases $G_{0,slot}$.
- Increasing $h$ lowers $f_{slot}$ and decreases $G_{0,slot}$.
- Increasing $\epsilon$ lowers $f_{slot}$ and increases $G_{0,slot}$.
- Increasing $U_w$ lowers $f_{slot}$ and decreases $G_{0,slot}$.
- Increasing $t_w$, $t_h$ and $d$ raises $f_{slot}$ somewhat.
- $U_h$ must be less than $L$ so the U-slot will fit on the patch when the two geometries are combined.

*C. Full U-slot Geometry Analysis and Iteration*

The two uncoupled resonator geometries are combined into a single structure upon which both CMA and the driven full-wave solve are computed. CMA will show in-phase and anti-phase coupled modes resonant according to (9). The driven full-wave impedance locus will have a loop, although it may not optimally reside within the RL limit circle on the Smith chart. From here, the geometry may be refined with these guidelines:

- Increasing the patch $Q$ or coupling factor $\kappa$ enlarges the Smith chart impedance locus loop, thus:
  o Decreasing $h$ enlarges the impedance locus loop.
  o Decreasing $W$ enlarges the impedance loop.
  o Increasing $U_w$ enlarges the impedance loop and moves it up (more inductive) on the Smith chart.
- Increasing $L$ enlarges the impedance loop and moves it down (more capacitive) on the Smith chart.
- Increasing $U_h$ shrinks the impedance loop and moves it up (more inductive) on the Smith chart.
- Increasing $p_0$ shrinks the impedance loop and moves it up (more inductive) on the Smith chart, if the probe is not very near the patch center.
- Increasing $t_w$ and $t_h$ or $d$ moves the impedance loop down (more capacitive) somewhat on the Smith chart.
- Increasing $U_0$ shrinks the impedance loop slightly.

VI. DESIGN EXAMPLE

The methodology is illustrated via design of a 2.4 GHz U-slot patch on a $h = 10$ mm PTFE substrate (modeled as permittivity $\epsilon_r = 2.1$). Conductors and dielectrics are modeled as ideal and thus the calculated radiation efficiency is 100%; small losses may be treated as a perturbation. FEKO [21] allows CMA with planar layered dielectric Green's function. We seek a 30% fractional 10 dB RL BW.

*A. Bandwidth-Optimal Coupled Resonances*

From (11), 30% BW implies $\kappa \sim 0.3$. Recall for $Z_0 = 50\ \Omega$ and 10 dB RL, $y_{\text{opt}} = 2.25$; thus, according to (11), $Q \sim 7.5$. Using (3) in (1) predicts $f_- = 2.04$ GHz and $f_+ = 2.76$ GHz.

*B. Designing the Uncoupled Resonators*

*1) Uncoupled Patch*

We seek $f_{\text{patch}} = 2.4$ GHz; closed-form formulae [33] yield $L = 34$ mm and we chose $W = 2L$ and $p_1 = L/2$. Although the probe has little effect in this geometry, we model it with arbitrary square cross section of 1 mm². CMA shows the $\text{TM}_{01}$ mode is resonant at 2.46 GHz. From (14) of the Appendix, $Q = 4.4$—less than the maximum. We adjust $L = 35$ mm; now $f_{\text{patch}} = 2.41$ GHz.

*2) Uncoupled Slot*

We seek $f_{\text{slot}} = 2.4$ GHz and $G_{0,\text{slot}} \sim 40 - 50$ mS. Assume $h$ and $\epsilon$ are fixed in this design; we may adjust only $U_l$ and $p_o$. The geometry of Fig. 8 is set with $U_l = \lambda_g/2$ where $\lambda_g = \lambda_0/\sqrt{\epsilon_{\text{eff}}}$ and $\epsilon_{\text{eff}} \sim (1 + \epsilon_r)/2$ [34]; thus $U_l = 50$ mm. We set $U_w \sim \kappa W = 20$ mm and center the probe via $p_o = (U_h - t_w)/2 \sim 7$ mm. We set $t_w = t_h = U_l/20 = 2.5$ mm arbitrarily and the probe cross-section as before; $t_w$, $t_h$ and $d$ can be used for fine-tuning later. CMA shows $f_{\text{slot}}$ is too high; $U_h$ is increased to 28 mm; now $f_{\text{slot}} = 2.41$ GHz, $G_{0,\text{slot}} = 43$ mS and $Q = 8.3$.

*3) Uncoupled Patch with Final Probe Location*

We return to the uncoupled patch and locate the probe as if the U-slot were centered in the patch, i.e., $U_0 = (L - U_h)/2 = 3.5$ mm; thus $p_1 \sim U_o + U_h - t_w - p_0 = 22$ mm. CMA yields $f_{\text{patch}} = 2.48$ GHz, $G_{0,\text{patch}} = 385$ µS and $Q = 4.3$.

*C. Full U-slot Geometry Analysis and Iteration*

The uncoupled patch and slot geometries are combined with the U-slot centered in the patch ($U_o = 3.5$ mm). We perform both CMA and the full-wave driven solve and identify CMA modes 1 and 3 as the in-phase and anti-phase modes, respectively, as shown in Fig. 15; Fig. 16 shows these modes are resonant at 2.00 GHz and 2.72 GHz—within 2% of that predicted by (1) and (3) with $\kappa = 0.3$.

Fig. 16 also shows slight interaction between the eigenvalues of the coupled modes (indicated by a minima in $|\lambda_3 - \lambda_1|$ near $f_0$) which is not present in the eigenvalue spectrum of Fig. 3(a); characteristic mode eigenvalue interaction has been associated with coupled mode theory [35]. However, unlike the examples of [35], the in-phase and anti-phase modes do not exchange characters during the interaction. We also note eigenvalue interaction is also determined by geometric symmetry [36] and inter-modal energy terms $\chi_{ij}$ [37]. It is plausible that the double symmetry of the U-slot geometry precludes eigenvalue interaction but that the presence of dielectric in the design example alters the inter-modal energy such that slight eigenvalue interaction is evident; further study of this topic is warranted.

The full-wave admittance is shown in Fig. 17 along with that of modes 1 and 3 as well as their parallel combination. The 10 dB RL BW is 31% and $\kappa = 0.30$ as calculated via (2). In this case, the initial combined geometry meets the stated design goals; if it had not, a few numerical iterations using the guidelines of Section V can quickly refine the performance. The Fig. 11(b) circuit model admittance is also shown in Fig. 17; the uncoupled resonator CMA data and resulting RLC values calculated via (6) are in Table III.

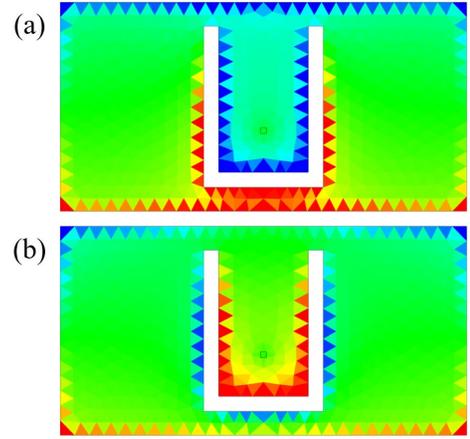

Figure 15. Characteristic charge distributions of (a) mode 1 and (b) mode 3 of the PTFE design example show (a) in-phase and (b) anti-phase relationships.

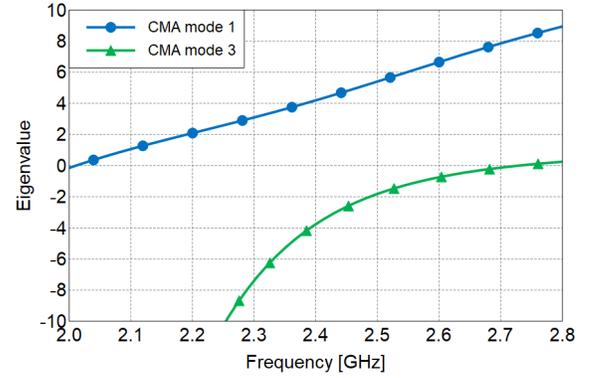

Figure 16. CMA mode 1 and 3 the in-phase and anti-phase modes, respectively) eigenvalues of the PTFE design example show resonance at 2.0 and 2.7 GHz—within 2% of that predicted by Coupled Mode Theory. Slight eigenvalue interaction is evident.

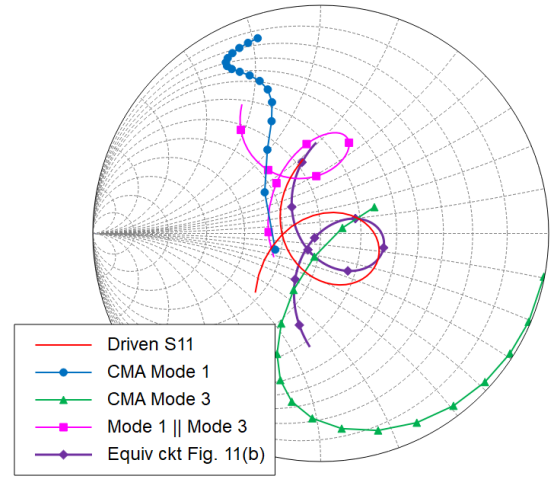

Figure 17. PTFE U-slot patch reflection coefficients for the full structure, CMA modes 1 & 3 (the in-phase and anti-phase modes, respectively), the parallel combination of CMA modes 1 & 3, and the Fig. 11(b) equivalent circuit with values from Table III.

TABLE III
UNCOUPLED RESONATOR CMA DATA AND RESULTING
EQUIVALENT CIRCUIT VALUES FOR PTFE DESIGN EXAMPLE

| Patch resonator | | Slot resonator | |
|---|---|---|---|
| $f_{\text{patch}}$ | 2.48 GHz | $f_{\text{slot}}$ | 2.41 GHz |
| $G_{0,\text{patch}}$ | 385 μS | $G_{0,\text{slot}}$ | 43 mS |
| $Q_{\text{patch}}$ | 4.3 | $Q_{\text{slot}}$ | 8.3 |
| $Q'$ | 6.0 | $Q'$ | 6.0 |
| $R_{\text{hp,patch}}$ | 93.5 kΩ | $R_{\text{hp,slot}}$ | 837 Ω |
| $L_{\text{patch}}$ | 1 μH | $L_{\text{slot}}$ | 9.21 nH |
| $C_{\text{patch}}$ | 4.12 fF | $C_{\text{slot}}$ | 0.473 pF |
| $k$ | 0.30 | $k$ | 0.30 |

The final dimensions (in mm) are: $h = 10$, $W = 68$, $L = 35$, $U_w = 20$, $U_h = 28$, $U_o = 3.5$, $p_o = 7$, $t_w = t_h = 2.5$, $d = 1$.

## VII. CONCLUSION

A first-principles mechanism of operation for the U-slot patch based on CMT, as revealed by CMA, has been presented. CMT was shown to be relevant in three independent ways: 1) by the presence of in-phase/anti-phase charge distributions (Fig. 6(c) and 6(d)), 2) by the ability of (1) to quantitatively describe the observed CMA frequency-splitting (demonstrated both in Fig. 7 as well as in the PTFE design example), and 3) via the success of a circuit model that explicitly shows coupling between two resonators (Fig. 11(b)) and is governed by CMT. A design methodology based on CMT was given; a key result is that the fractional bandwidth, normalized coupling coefficient, and ratio of dimensions $U_w/W$ are approximately equal (as captured by (4) and (11)). Finally, we believe the general concepts and methods presented here are applicable to similar wideband antenna geometries such as probe-fed patches with "V"- or "W"-shaped slots and the "E"-shaped patch.

## APPENDIX

CMA is a modal decomposition based on MoM [38], [39], [40]; reviews are presented in [41] and [42]. Within CMA, a set of real orthogonal basis currents $J_n$ result from an eigenvalue equation of the MoM impedance matrix $[Z] = [R] + j[X]$:

$$[X] J_n = \lambda_n [R] J_n \qquad (12)$$

where $\lambda_n$ is the eigenvalue. Modes are orthogonal and traditionally normalized such that they radiate unit power, i.e., $\langle J_m^*, R J_n \rangle = \delta_{mn}$, where $\delta_{mn}$ is the Kronecker delta [39]. Near- and far-field distributions are associated with each mode. Harrington showed [39]:

$$2\omega (W_m - W_e) = \lambda_n \qquad (13)$$

where $W_m$ and $W_e$ are the modal magnetic and electric energies, respectively; at resonance $\lambda_n = 0$. The modal quality factor $Q$ is calculated as [31], [43], [44]:

$$Q = \frac{\omega}{2} \frac{d\lambda_n}{d\omega}\bigg|_{\lambda_n=0}. \qquad (14)$$

If a gap voltage source is present, the admittance $Y[m]$ at an unknown $m$ can be calculated as a sum of modal admittances at [38], [39], [45]:

$$Y[m] = \sum_n \frac{J_n[m]^2}{1 + \lambda_n^2} (1 - j\lambda_n), \qquad (15)$$

which implies that the equivalent circuit for a structure is the parallel combination of individual modal circuits (which themselves are series resonances).

The extent to which a mode is excited by a source $E_{\text{tan}}^{\text{i}}$ is quantified by the modal weighting coefficient [39]:

$$\alpha_n = \frac{\langle J_n, E_{\text{tan}}^{\text{i}} \rangle}{1 + j\lambda_n}, \qquad (16)$$

with which the total current distribution on the structure can be written as $J_{\text{total}} = \sum_n \alpha_n J_n$.

Recently, [35] connected CMA and CMT through the so-called "eigenvalue crossing avoidance" phenomenon, which was shown to be governed by a relation similar to (1).

## ACKNOWLEDGMENT

The authors thank Dylan A. Crocker, Thomas E. Roth and Jeffery T. Williams of Sandia National Laboratories for providing helpful criticism of the manuscript.

## REFERENCES

[1] T. Huynh and K. F. Lee, "Single-layer single-patch wideband microstrip antenna," in Electronics Letters, vol. 31, no. 16, pp. 1310-1312, 3 Aug. 1995.
[2] K. F. Lee, K. M. Luk, K. F. Tong, S. M. Shum, T. Huynh and R. Q. Lee, "Experimental and simulation studies of the coaxially fed U-slot rectangular patch antenna," in IEE Proceedings - Microwaves, Antennas and Propagation, vol. 144, no. 5, pp. 354-358, Oct. 1997.
[3] K. F. Lee, S. L. S. Yang and A. Kishk, "The versatile U-slot patch antenna," 2009 3rd European Conference on Antennas and Propagation, Berlin, 2009, pp. 3312-3314.
[4] M. Clenet and L. Shafai, "Multiple resonances and polarisation of U-slot patch antenna," in Electronics Letters, vol. 35, no. 2, pp. 101-103, 21 Jan. 1999.
[5] K.-F. Tong, K.-M. Luk, K.F. Lee and R.Q. Lee, "A broad-band U-slot rectangular patch antenna on a microwave substrate," in IEEE Transactions on Antennas and Propagation, vol. 48, no. 6, pp. 954-960, June 2000.
[6] Y. L. Chow, Z. N. Chen, K. F. Lee and K. M. Luk, "A design theory on broadband patch antennas with slot," IEEE Antennas and Propagation Society International Symposium. 1998 Digest, Atlanta, GA, 1998, pp. 1124-1127 vol.2.
[7] R. Bhalla and L. Shafai, "Resonance behavior of single U-slot and dual U-slot antenna," IEEE Antennas and Propagation Society International Symposium. 2001 Digest, Boston, MA, USA, 2001, pp. 700-703 vol.2.
[8] S. Weigand, G. H. Huff, K. H. Pan and J. T. Bernhard, "Analysis and design of broad-band single-layer rectangular U-slot microstrip patch antennas," in IEEE Transactions on Antennas and Propagation, vol. 51, no. 3, pp. 457-468, March 2003.
[9] V. Natarajan and D. Chatterjee, "An Empirical Approach for Design of Wideband, Probe-Fed, U-Slot Microstrip Patch Antennas on Single-layer, Infinite, Grounded Substrates," ACES Journal, vol. 18, no. 3, November 2003.

[10] J. A. Ansari and R. Brij Ram, "Analysis of broad band U-slot microstrip patch antenna," Microw. Opt. Technol. Lett., vol. 50, pp. 1069-1073, 2008.

[11] Y. Chen and C. Wang, "Characteristic-Mode-Based Improvement of Circularly Polarized U-Slot and E-Shaped Patch Antennas," in IEEE Antennas and Wireless Propagation Letters, vol. 11, pp. 1474-1477, 2012.

[12] M. Capek, P. Hazdra, P. Hamouz, and J. Eichler, "A Method for Tracking Characteristic Numbers and Vectors," in Progress in Electromagnetics Research B, Vol. 33, pp. 115-134, 2011.

[13] N. M. Mohamed-Hicho, E. Antonino-Daviu, M. Cabedo-Fabrés, J. P. Ciafardini and M. Ferrando-Bataller, "On the interaction of Characteristic Modes in slot antennas etched on finite ground planes," 2016 10th European Conference on Antennas and Propagation (EuCAP), Davos, 2016, pp. 1-5.

[14] Z. Yang, D. Su, Y. Li and Y. Liu, "An improved method for tracking of Characteristic Modes," 2016 IEEE International Conference on Computational Electromagnetics (ICCEM), Guangzhou, 2016, pp. 103-105.

[15] M. M. Elsewe and D. Chatterjee, "Modal analysis of patch slot designs in microstrip patch antennas," 2016 IEEE/ACES International Conference on Wireless Information Technology and Systems (ICWITS) and Applied Computational Electromagnetics (ACES), Honolulu, HI, 2016, pp. 1-2.

[16] M. Khan and D. Chatterjee, "Characteristic Mode Analysis of a Class of Empirical Design Techniques for Probe-Fed, U-Slot Microstrip Patch Antennas," in IEEE Transactions on Antennas and Propagation, vol. 64, no. 7, pp. 2758-2770, July 2016.

[17] M. Khan and D. Chatterjee, "Characteristic modes for U-slot's feed placement," 2017 IEEE International Symposium on Antennas and Propagation & USNC/URSI National Radio Science Meeting, San Diego, CA, 2017, pp. 743-744.

[18] M. Khan and D. Chatterjee, "Analysis of Reactive Loading in a U-Slot Microstrip Patch Using the Theory of Characteristic Modes," in IEEE Antennas and Propagation Magazine, vol. 60, no. 6, pp. 88-97, Dec. 2018.

[19] T. C. LaPointe, "Characterization of Wideband U-slot Patch Antennas Through Characteristic Modal Analysis and Coupled Mode Theory," M.S. thesis, Dept. ECE, Univ. New Mexico, Albuquerque, NM, 2018.

[20] J. Borchardt and T. LaPointe, "Analysis of a U-slot Patch Using Characteristic Mode Analysis and Coupled Mode Theory," 2019 IEEE International Symposium on Antennas and Propagation and USNC-URSI Radio Science Meeting, Atlanta, GA July 7-12, 2019. To be published.

[21] Altair Engineering, Inc. FEKO [Online] Available: https://altairhyperworks.com/product/FEKO

[22] J. R. Pierce "Coupled modes," in Almost All About Waves, MIT Press 1974, ch. 6 p. 47.

[23] H. A. Haus and W. Huang, "Coupled-mode theory," in Proceedings of the IEEE, vol. 79, no. 10, pp. 1505-1518, Oct. 1991.

[24] S. L. Chuang, "Waveguide couplers and coupled mode theory" in Physics of Optoelectronic Devices, 1st ed., New York, NY, USA: Wiley, 1995, ch. 8, sec. 2.2, p. 291.

[25] J. Hong, "Couplings of asynchronously tuned coupled microwave resonators," in IEE Proceedings - Microwaves, Antennas and Propagation, vol. 147, no. 5, pp. 354-358, Oct. 2000.

[26] S. Bhardwaj and Y. Rahmat-Samii "A comparative study of c-shaped, e-shaped, and u-slotted patch antennas," in Microw. Opt. Technol. Lett., vol. 54, pp. 1746-1757, 2012.

[27] A. Deshmukh and K. P. Ray, "Analysis of Broadband Variations of U-slot cut Rectangular Microstrip Antennas," in IEEE Antennas and Propagation Magazine, vol. 57, no. 2, pp. 181-193, April 2015.

[28] H. Xu, D. R. Jackson and J. T. Williams, "Comparison of models for the probe inductance for a parallel-plate waveguide and a microstrip patch," in IEEE Transactions on Antennas and Propagation, vol. 53, no. 10, pp. 3229-3235, Oct. 2005.

[29] J. J. Lee, "Slotline impedance," in IEEE Transactions on Microwave Theory and Techniques, vol. 39, no. 4, pp. 666-672, April 1991.

[30] D.M. Pozar, "Parallel Resonant Circuit" in Microwave Engineering, $4^{th}$ ed., John Wiley & Sons, 2011, ch. 6, p. 275.

[31] J. J. Adams and J. T. Bernhard, "Broadband Equivalent Circuit Models for Antenna Impedances and Fields Using Characteristic Modes," IEEE Transactions on Antennas and Propagation, vol. 61, no. 8, pp. 3985-3994, Aug. 2013.

[32] D. Kajfez, "Dual resonance," IEE Proceedings H - Microwaves, Antennas and Propagation, vol. 135, no. 2, pp. 141-143, April 1988.

[33] Antenna Engineering Handbook, 4th ed., McGraw-Hill, New York, NY, USA, 2007, ch. 7.

[34] K. C. Gupta, R. Garg, I. J. Bahl, "Slotlines I" in Microstrip Lines and Slotlines, 1st ed., Dedham, MA, USA: Artech House, 1979, ch. 5, p. 199.

[35] K. R. Schab, J. M. Outwater, M. W. Young and J. T. Bernhard, "Eigenvalue Crossing Avoidance in Characteristic Modes," in IEEE Transactions on Antennas and Propagation, vol. 64, no. 7, pp. 2617-2627, July 2016.

[36] K. R. Schab and J. T. Bernhard, "A Group Theory Rule for Predicting Eigenvalue Crossings in Characteristic Mode Analyses," in IEEE Antennas and Wireless Propagation Letters, vol. 16, pp. 944-947, 2017.

[37] K. R. Schab, J. M. Outwater and J. T. Bernhard, "Classifying characteristic mode crossing avoidances with symmetry and energy coupling," 2016 IEEE International Symposium on Antennas and Propagation (APSURSI), Fajardo, 2016, pp. 13-14.

[38] R. Garbacz and R. Turpin, "A generalized expansion for radiated and scattered fields," in IEEE Transactions on Antennas and Propagation, vol. 19, no. 3, pp. 348-358, May 1971.

[39] R. Harrington and J. Mautz, "Theory of characteristic modes for conducting bodies," in IEEE Transactions on Antennas and Propagation, vol. 19, no. 5, pp. 622-628, September 1971.

[40] R. Harrington and J. Mautz, "Computation of characteristic modes for conducting bodies," in IEEE Transactions on Antennas and Propagation, vol. 19, no. 5, pp. 629-639, September 1971.

[41] M. Cabedo-Fabres, E. Antonino-Daviu, A. Valero-Nogueira and M. F. Bataller, "The Theory of Characteristic Modes Revisited: A Contribution to the Design of Antennas for Modern Applications," in IEEE Antennas and Propagation Magazine, vol. 49, no. 5, pp. 52-68, Oct. 2007.

[42] M. Vogel, G. Gampala, D. Ludick, U. Jakobus and C. J. Reddy, "Characteristic Mode Analysis: Putting Physics back into Simulation," in IEEE Antennas and Propagation Magazine, vol. 57, no. 2, pp. 307-317, April 2015.

[43] M. Cabedo-Fabres, "Systematic design of antennas using the theory of characteristic modes," Ph.D. dissertation, Universidad Politecnica de Valencia, Valencia, Spain, 2007, p. 35.

[44] R. Harrington and J. Mautz, "Control of radar scattering by reactive loading," in IEEE Transactions on Antennas and Propagation, vol. 20, no. 4, pp. 446-454, July 1972.

[45] A. Yee and R. Garbacz, "Self- and mutual-admittances of wire antennas in terms of characteristic modes," in IEEE Transactions on Antennas and Propagation, vol. 21, no. 6, pp. 868-871, November 1973.